\documentclass[prl,aps,twocolumn,amsmath,showpacs,amssymb,letterpaper]{revtex4}
\usepackage{hyperref}
\pdfoutput=1

\usepackage{graphic x}
\usepackage{here}

\def\cal#1{\mathcal{#1}}
\def\beq{\begin{equation}}
\def\eeq{\end{equation}}
\def\bea{\begin{eqnarray}}
\def\eea{\end{eqnarray}}

\def\dc{^{\circ}{\rm C}}

\def\kt{k_{\rm B} T}

\begin{document}
\title{Stretching chimeric DNA: a test for the putative S-form}
\author{Stephen Whitelam$^{1,2}$} 
\author{Sander Pronk$^2$} 
\author{Phillip L. Geissler$^2$}
\affiliation{$^1$Systems Biology Centre, University of Warwick, Coventry CV4 7AL, UK \\
$^2$Department of Chemistry, University of California at Berkeley, and Physical Biosciences and Materials Sciences Divisions, Lawrence Berkeley National Laboratory, Berkeley, CA 94720}
\begin{abstract}
Double-stranded DNA `overstretches' at a pulling force of about 65 pN, increasing in length by a factor of 1.7. The nature of the overstretched state is unknown, despite its considerable importance for DNA's biological function and technological application. Overstretching is thought by some to be a force-induced denaturation, and by others to consist of a transition to an elongated, hybridized state called S-DNA. Within a statistical mechanical model we consider the effect upon overstretching of extreme sequence heterogeneity. `Chimeric' sequences possessing halves of markedly different AT composition elongate under fixed external conditions via distinct, spatially segregated transitions. The corresponding force-extension data display two plateaux at forces whose difference varies with pulling rate in a manner that depends qualitatively upon whether the hybridized S-form is accessible. This observation implies a test for S-DNA that could be performed in experiment. Our results suggest that qualitatively different, spatially segregated conformational transitions can occur at a single thermodynamic state within single molecules of DNA. 
\end{abstract}
\maketitle

\section{Introduction}
Double-stranded DNA elongates abruptly at a force of about 65 pN if it is pulled along its axis~\cite{ten_years,ritort}. The resulting `overstretched' form of the molecule is approximately 1.7 times longer than helical B-DNA. Overstretching is of crucial importance for the biological function of DNA: the bacterial protein RecA elongates DNA by a factor of 1.5 upon binding~\cite{reca1,reca2,reca3}, a mechanism central to homologous recombination and to chromosomal segregation during cell division~\cite{bio_textbook}. However, the nature of the overstretched state remains a source of considerable controversy: some think overstretched DNA {\em in vitro} to be a hybridized form called S-DNA (the `B-to-S' picture)~\cite{bustamante, cocco,s_form1, s_form2, s_form3,zhou}, while a competing picture considers overstretching to signal a conversion to unhybrizided single strands (the `force-melting' picture)~\cite{bloomfield0,bloomfield1,bloomfield2,force_melting, williams, melting_prl}. We therefore lack full understanding of the basic mechanical and thermodynamic properties of a molecule of central importance to biology and of rapidly increasing importance to technology~\cite{linkers,cond,golden,handshake}.

It is not possible to conclusively validate or rule out either picture of overstretching on the basis of existing thermodynamic force-extension data. The variation of overstretching force with parameters such as temperature and salt concentration~\cite{bloomfield0,bloomfield1,bloomfield2}, both of which are known to change the melting properties of DNA, implies that melting of the double helix plays an important role in its elongation. Combining these data with the observation that 65 pN of tension provides just enough mechanical energy to render B-DNA single-stranded or `molten'~\cite{bloomfield1} strongly suggests that overstretching {\em is} melting, induced by a pulling force. However, some authors~\cite{cocco} charge that molten DNA, in the sense of two parallel but noninteracting single strands, is thermodynamically unstable to a form of DNA in which one strand has frayed or `unpeeled' from the other, and no longer bears tension. The latter state may arise when melting occurs in the vicinity of `nicks' in the phosphate backbone. Nicks are present in many overstretching experiments (see, however, Ref.~\cite{mara}), and are considered to relieve the torsional constrant on the structure, so permitting the two strands to rotate freely when stretched. The instability of melting to unpeeling implies that for nicked DNA the noninteracting parallel-strand conformation is not the stable state at 65 pN. However, this stable state appears not to be the unpeeled form of the molecule, either. Under some conditions overstretched DNA possesses a mechanical stiffness far in excess of that of single-stranded forms of the molecule~\cite{cocco}, implying that the stable overstretched form at 65 pN is double-stranded, as in the S-DNA picture. This view is bolstered by the observation that a transition to single-stranded (unpeeled) DNA is sometimes observed only at forces well in excess of 100 pN~\cite{rief,cs}. But these observations pose a problem: assuming that overstretching involves the interconversion of two hybridized states, which presumably differ principally in their enthalpic rather than their entropic properties, how does one rationalize the striking temperature dependence of overstretching?
 
Just as thermodynamic data fail to distinguish between force-melting and B-to-S pictures, so current imaging techniques cannot resolve the structure of overstretched DNA. Further, atomistic simulations of overstretching~\cite{harris,md1,md2}, while offering valuable insight into molecular mechanisms, cannot approach (by orders of magnitude) the length and time scales characteristic of experiment. We argue, by contrast, that the {\em kinetics} of force-extension data, combined with predictions from coarse-grained, statistical mechanical models, may offer a means of discriminating between the two scenarios. 

In typical experiments, an optical tweezers~\cite{tweezers1,tweezers2} or atomic force microscope (AFM) is used to overstretch a single molecule of DNA at constant rate of extension. The loading rate is then reversed, allowing the molecule to recover its original length. If this stretching-shortening cycle is carried out at low temperature, the force-extension traces for each stage superpose, indicating a reversible transition. As temperature increases, stretching and shortening traces become distinct, signaling hysteresis. The degree of this hysteresis increases as temperature increases~\cite{hanbin}. 

Such temperature-dependent kinetics provides a stringent test of theories of overstretching. Hysteresis is very unusual in a system whose extent is macroscopic in only one dimension, and whose fluctuations involve local free energy differences not much greater than the thermal energy, $k_{\rm B} T$. Well-known causes of hysteresis in physical systems include strong local interactions that lead to an unfavorable surface tension between coexisting phases within, for instance, magnetic materials or liquid-vapor phase transitions. Overcoming this surface tension in order to nucleate and grow domains of the thermodynamically dominant phase results in sluggish kinetics and hysteresis. However, hysteresis associated with strong interactions increases in degree as temperature {\em decreases}, by contrast with overstretching hysteresis. Moreover, surface tension does not grow with domain size in a quasi-one dimensional structure such as DNA. One is led to the conclusion that hysteresis in stretching data implies the emergence of concerted, long-wavelength correlations.

Long-wavelength correlations can emerge from the detachment and re-annealing of strands~\cite{bustamante,bloomfield1}. The authors of Ref.~\cite{cocco} demonstrated that a kinetic model of strand separation (unpeeling) displayed hysteresis similar in character to that seen in experiment. Building on these observations, we introduced a discrete statistical mechanical model designed to assess the kinetics associated with both the `B-to-S' and `force-melting' pictures of overstretching~\cite{us}. The model is resolved at the level of individual basepairs and assumes that DNA may locally adopt certain discrete conformational states. It is inspired by and borrows features from models of DNA undergoing thermal melting~\cite{BZ, poland,santa} and overstretching~\cite{bloomfield1,cocco}. Our conclusion is that indeed long-wavelength correlations of the kind propagated by separating and re-annealing strands induce hysteresis. Crucially, in order to observe a {\em progression} with temperature of the degree of hysteresis, we require a progression with temperature of the nature of the overstretched state. This progression results within our model from a competition between the basepairing energy of S-DNA and the entropy liberated upon unpeeling. At low temperatures, S-DNA predominates as the overstretched form. Our model of the B-to-S transition involves only local free energy barriers of magnitude $\leq 10 \, \kt$, and at pulling rates considered occurs in equilibrium. At high temperature, unpeeling occurs, whose associated long-wavelength correlations give rise to hysteresis. 
\begin{figure}[h!] 
\label{}
\centering
\includegraphics[width=\linewidth]{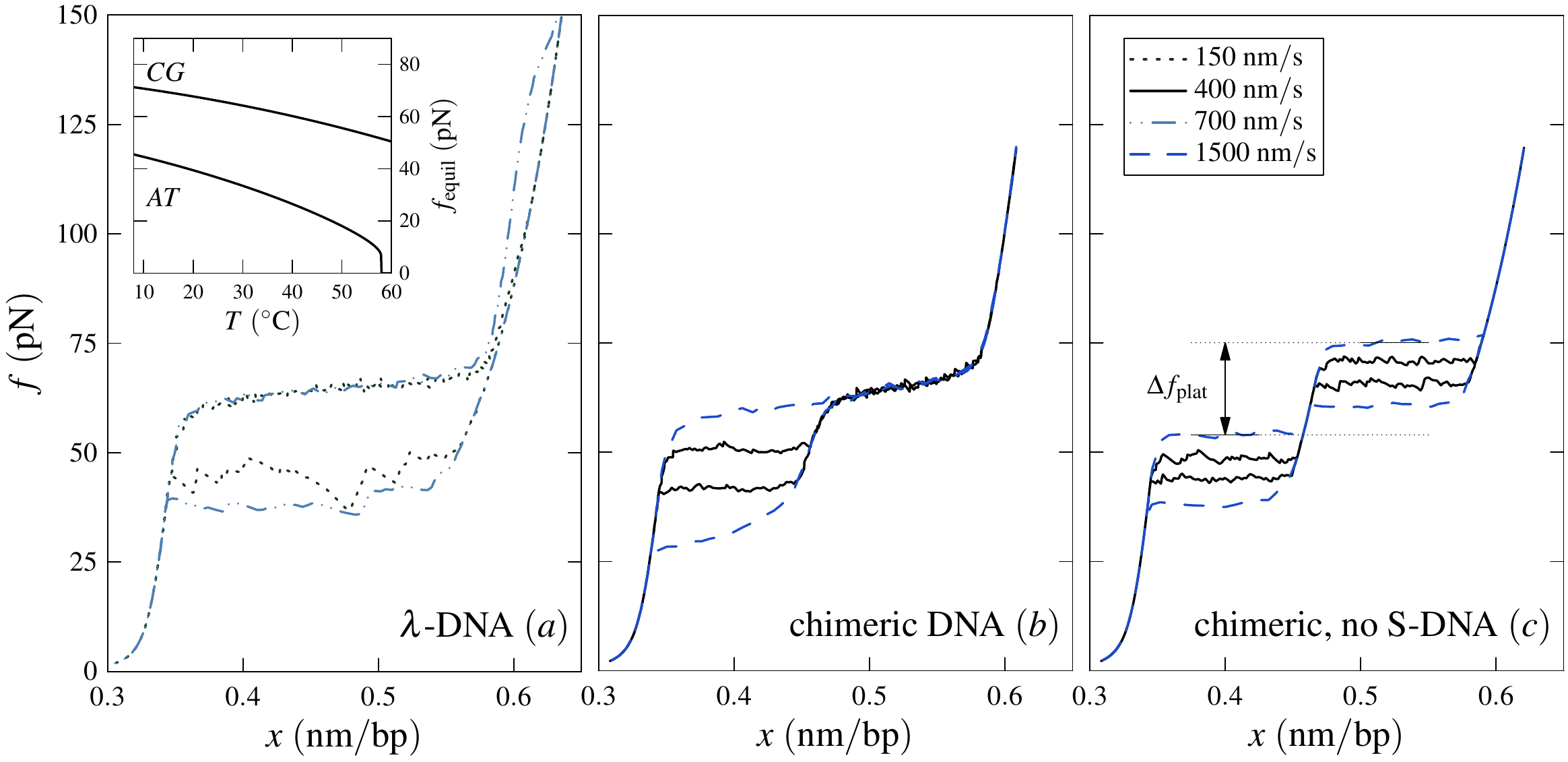} 
 \caption{Overstretching natural and designed sequences. (a) Simulated $\lambda$-DNA (40 kbp, 150 mM NaCl, $21 \dc$, pullrates 150 and 700 nm/s) displays a B-to-S transition at 65 pN followed at higher forces by a rate-dependent unpeeling. (b,c) By contrast, simulations of strongly heterogeneous (chimeric) sequences (10 kbp, 150 mM NaCl, 16$\dc$, $d=0$, pullrates 400 and 1500 nm/s) exhibit both transitions in a spatially segregated fashion. Corresponding force-extension data display double plateaux, whose forces vary strongly with pulling rate if plateaux correspond to unpeeling, and are insensitive to pulling rate if plateaux signal B-to-S conversion. When S-DNA is included in the model, the AT-rich half can unpeel (b, lower plateau) and the CG-rich half can convert to S-DNA (b, upper plateau); when S-DNA is disallowed both halves unpeel (c). Inset: force as a function of temperature at which the basepair doublets CG:CG and AT:AT unpeel in preference to remaining as B-DNA. These unpeeling transitions compete with B-to-S conversion at 65 pN.}
 \end{figure}

Consequently, we interpret overstretching as a process involving competing conformational transitions, with elements of {\em both} pictures required to explain experimental results. This view has several qualitiative implications for experiment, such as the prediction that hysteresis should be a non-monotonic function of pulling rate~\cite{us}. Here we examine the consequences of this picture of overstretching in a new context, by exploring within our model the effect of extreme sequence heterogeneity upon the competition between unhybridized and hybridized elongated states. We are motivated by the expectation that appropriate sequence design could lead, within the same molecule, to distinct transitions from B-DNA to two {\em different} overstretched states. In stretching experiments at room temperature, bacterial $\lambda$-phage DNA ($\lambda$-DNA) (roughly 50:50 AT:CG content, without long-ranged sequence correlations~\footnote{Within $\lambda$-DNA, the probability of consecutive stretches of AT or CG basepairs of length $\ell$ decreases exponentially with $\ell$.}) displays a force-extension plateau at about 65 pN, independent of pulling rate between 150 and 3000 nm/s. A second transition is observed at higher forces, strongly dependent upon pulling rate~\cite{rief,cs}. This behavior is illustrated by model calculations in Figure 1, left panel. These observations have been rationalized as two, temporally segregated transitions, the first an equilibrium B-to-S conversion, the second signaling out-of-equilibrium unpeeling~\cite{cocco,us}. We demonstrate here that one can dramatically exaggerate the differences between these transitions by designing sequences that permit such changes to occur in a {\em spatially} segregated fashion. We suggest that testing for similar segregation in experiment will reveal if indeed there exists an elongated, hybridized state.

\section{Model} 
Our model resolves detail at the level of individual basepairs. We assume that each basepair, of type AT, TA, CG or GC (we assume fully complementary alignment), may instantaneously adopt one of four discrete states. These are the helical B-form; two different unhybridized forms (M, or molten, corresponding to internal molten bubbles; and U, or unpeeled, in which only one strand is load-bearing); and the putative S-form. The model takes as its input coarse-grained free energetic properties of each state, and via a Monte Carlo algorithm offers kinetic predictions on length- and timescales characteristic of experiments. Full details are given in Ref.~\cite{us}, with two modifications considered in this paper. The first accounts for a nearest-neighbor dependence of basepairing-stacking energies~\cite{santa}, while the second accounts for the sequence-dependence of the heat capacity of melting~\cite{bloomfield1,bloomfield2}. We assume that specific heats of melting of AT and CG basepairs are zero at (i.e. Taylor expanded about), respectively, the (salt-dependent) melting temperatures of AT- and CG-DNA. This gives a model in which the stabilities of AT and CG basepairs are more similar at lower temperatures or higher salt conditions than under the converse conditions. While such details matter greatly when considering changes in overstretching behavior with temperature, they play essentially no role in determining changes with pulling rate. We evolve our model according to a dynamic protocol designed to mimic optical trap or AFM pulling experiments. We increment at constant rate the position of one end of the molecule, allowing the position of the molecule end tethered to an imaginary optical trap or cantilever to fluctuate. We calculate the resulting tension subject to the constraint of mechanical equilibrium. Basepair fluctuations are assumed to occur on a timescale derived from fluorescence correlation spectroscopy~\cite{bubbles0,bubbles1,bubbles2}; we discuss this choice in Appendix B.
\begin{figure}[!h] 
\label{}
\centering
\includegraphics[width=0.8\linewidth]{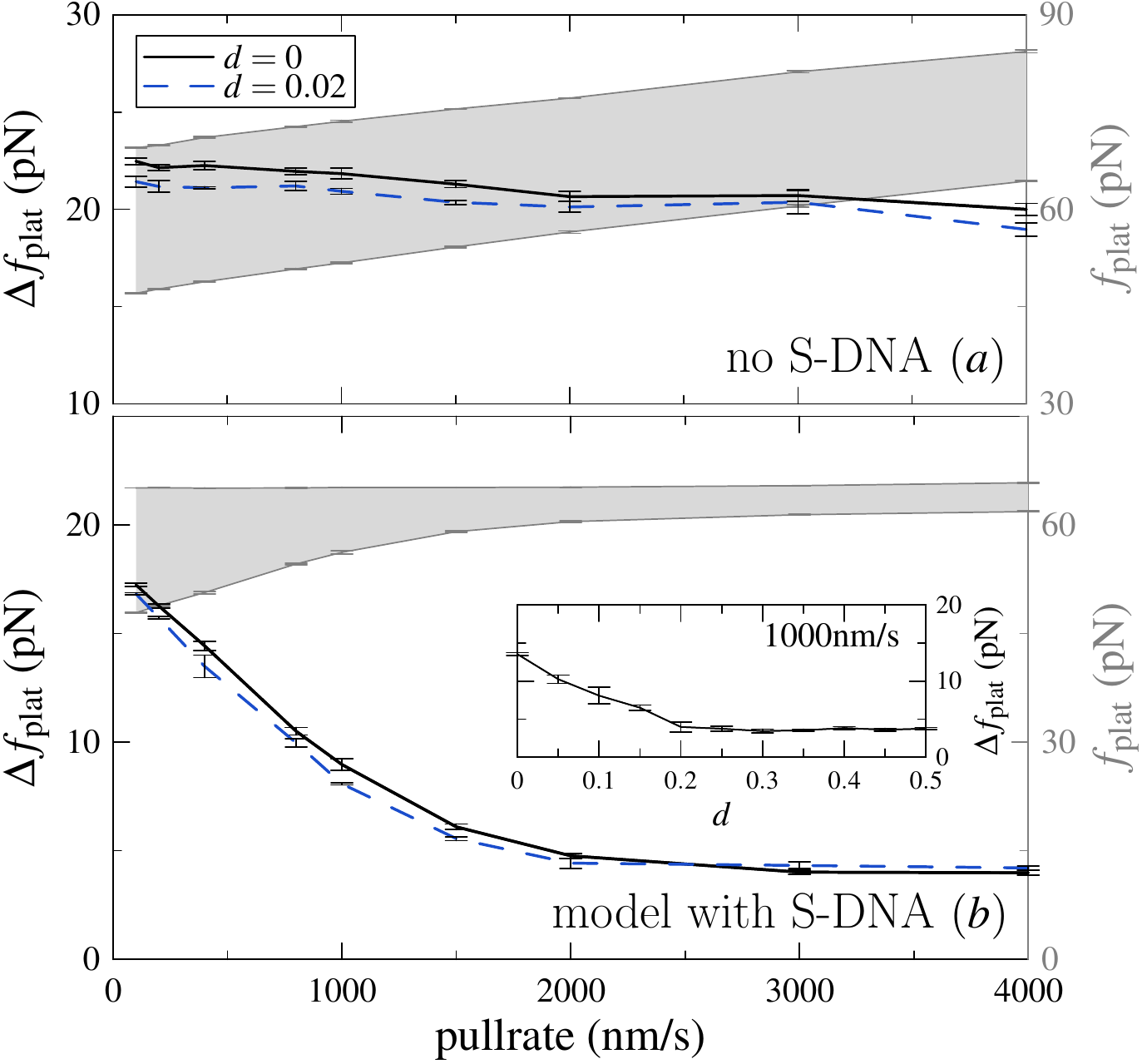} 
\caption{\label{fig2} Simulated force-extension data for strongly heterogeneous sequences bear a kinetic `fingerprint' of the nature of the spatially segregated overstretching mechanisms. The pulling rate-dependence of $\Delta f_{\rm plat}$ for 10 kbp chimeric sequences at $16^{\circ}$C and 150 mM NaCl ($d=0$ [each data point is the mean of 5 simulations]; and $d=0.02$ [each data point is the mean of 2 simulations]) is qualitatively altered by the presence of an elongated, hybridized state. Plateaux height difference if S-DNA is disallowed (a) is a weak function of pulling rate, but is a strong function of pulling rate if S-DNA is included in the model (b). Absolute plateaux heights border the shaded areas (main figure, right axis). Inset: $\Delta f_{\rm plat}$ at 1000 nm/s as a function of sequence disorder $d$ within the B-to-S model. The double plateaux structure (corresponding to nonzero $\Delta f_{\rm plat}$) is visible for $d \leq 0.15$. }
\end{figure}

\section{Results} 
We consider simulations of `chimeric' DNA fragments whose left halves consist of basepairs having chemical composition AT or TA with respective probabilities $(1-d)/2$, and CG or GC with respective probabilities $d/2$. We call $d$ the `sequence disorder' parameter. The right halves have a corresponding structure with the replacement $d \to 1-d$. Thus $d=0$ indicates a perfectly segregated sequence, while $d=1/2$ corresponds to a completely random sequence. We place one nick at the left extremity of the AT-rich half. The nick permits unpeeling, which at the pulling rates and temperatures we consider occurs in preference to the internal molten bubble configuration (M). We perform calculations on sequences of lengths from 10 kilobasepairs to 200 basepairs.

In Figure 1 (b,c) we show that in simulations modeling stretching near room temperature these constructs display two distinct force-extension plateaux, corresponding to unpeeling of the AT-rich half (b,c, lower plateaux) and one of two fates for the CG-rich half. When we permit the model to access the S-state, the CG half can elongate by way of the B-to-S transition (b, upper plateau). When S-DNA is not included in the model, the CG half unpeels, albeit at a higher force than does the AT-rich half (c, upper plateau). Plateaux resulting from unpeeling, which occurs out of equilibrium, display strong pulling rate dependence~\cite{cocco,rief,us}; plateaux signalling equilibrium B-to-S conversion do not. Consequently, the difference in `height' (force) between double plateaux, $\Delta f_{\rm plat}$, varies with pulling rate in a manner that depends qualitatively upon whether the molecule may access the hybridized S-state. When S-DNA is accessible, $\Delta f_{\rm plat}$ varies strongly with pulling rate. When S-DNA is inaccessible we find that the difference in height between unpeeling plateaux is almost insensitive to pulling rate, because the mechanism of strand separation is similarly hysteretic for both halves.  We note that at higher forces, within the B-to-S model, one can observe a three-stage transition within chimeric sequences (Appendix A). We note also that at temperatures high enough that S-DNA is unstable at about $65$ pN to unpeeling of CG basepairs, the B-to-S model describes a variation of $\Delta f_{\rm plat}$ with pulling rate similar to that of the force-melting model (data not shown).
\begin{figure}[!h] 
\label{}
\centering
\includegraphics[width=0.8\linewidth]{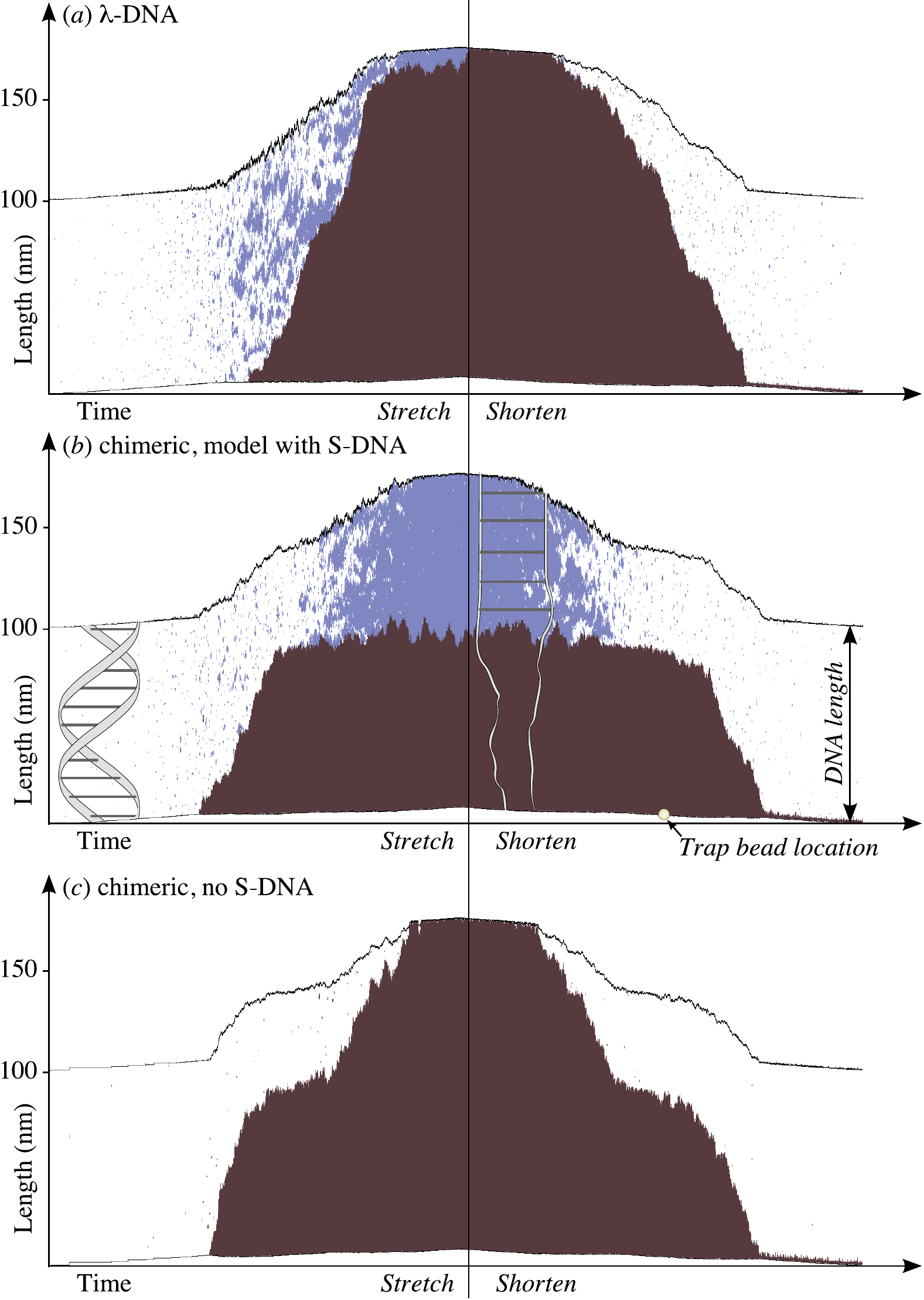} 
\caption{\label{figtraj} Microscopic configurations as a function of time from simulations of 300-basepair fragments of $\lambda$-DNA (a) and chimeric DNA (b,c) at $21^{\circ}$C and 150 mM NaCl, for pulling rate 1000 nm/s. White indicates B-DNA, red indicates unhybridized DNA, and blue indicates S-DNA. The spatially distinct transitions observed within chimeric sequences are of different character depending on whether S-DNA is assumed to exist (b) or not (c). The spatial scale indicates molecular length, with the extension modeling the departure of the `trapped' bead from the trap center reduced by a factor of $10$ for clarity. The bottom (top) halves of chimeric sequences are AT (CG) rich. The maximum force attained is 90 pN.}
\end{figure}

In Figure~\ref{fig2} we quantify these observations by plotting $\Delta f_{\rm plat}$ versus pulling rate with S-DNA included in the model (a) or not (b). The behavior of $\Delta f_{\rm plat}$ is qualitatively and strikingly different within the two scenarios, and constitutes the key result of this paper. The corresponding pulling experiments would reveal, firstly, if force-extension data for chimeric sequences exhibit double plateaux, and secondly, if so, how these plateaux vary with pulling rate. We argue that pronounced variation with pulling rate of $\Delta f_{\rm plat}$ would imply overstretching via two distinct mechanisms, suggesting that elongation is possible via both unhybridized and hybridized states. 

In Figure~\ref{figtraj} we show the time-dependent microscopic configurations that underlie these force-extension data, comparing $\lambda$-DNA (a) with chimeric DNA (b,c). The spatially distinct transitions observed within strongly heterogeneous sequences are of different character depending on whether S-DNA is assumed to exist (b) or not (c). The microscopic signature of unpeeling is a long-wavelength drift of the domain wall separating hybridized and unhybridized conformations.

Our results indicate that clean segregation between overstretching transitions (signaled by distinct double plateaux in force-extension data) occurs for molecules as short as 200 bp (data not shown), and for sequence disorder parameters $d$ as large as $d \approx 0.15$. For the B-to-S model, these segregated transitions are different in nature (unpeeling versus B-to-S conversion) for temperatures $T \leq 25^{\circ}$C at 150 mM NaCl (and for higher temperatures at higher salt concentrations). These results suggest that DNA-elongating proteins, such as RecA, could induce within DNA stretching transitions whose character differs with sequence composition, conferring upon the elongated RecA-DNA complex~\cite{reca1,reca2} a sequence-dependent elasticity. Our results also suggest that precise melting of localized regions of DNA can be effected by stretching molecules whose sequences are appropriately designed. Such precision de-hybridization would facilitate basepair doping with ligands, allowing one to modify the conductance properties of the molecule in a spatially heterogeneous fashion. Control of conductance would enhance DNA's usefulness as a nanowire in molecular electronics devices.
      \begin{figure}[!h] 
   \label{}
   \centering
         \includegraphics[width=0.7 \linewidth]{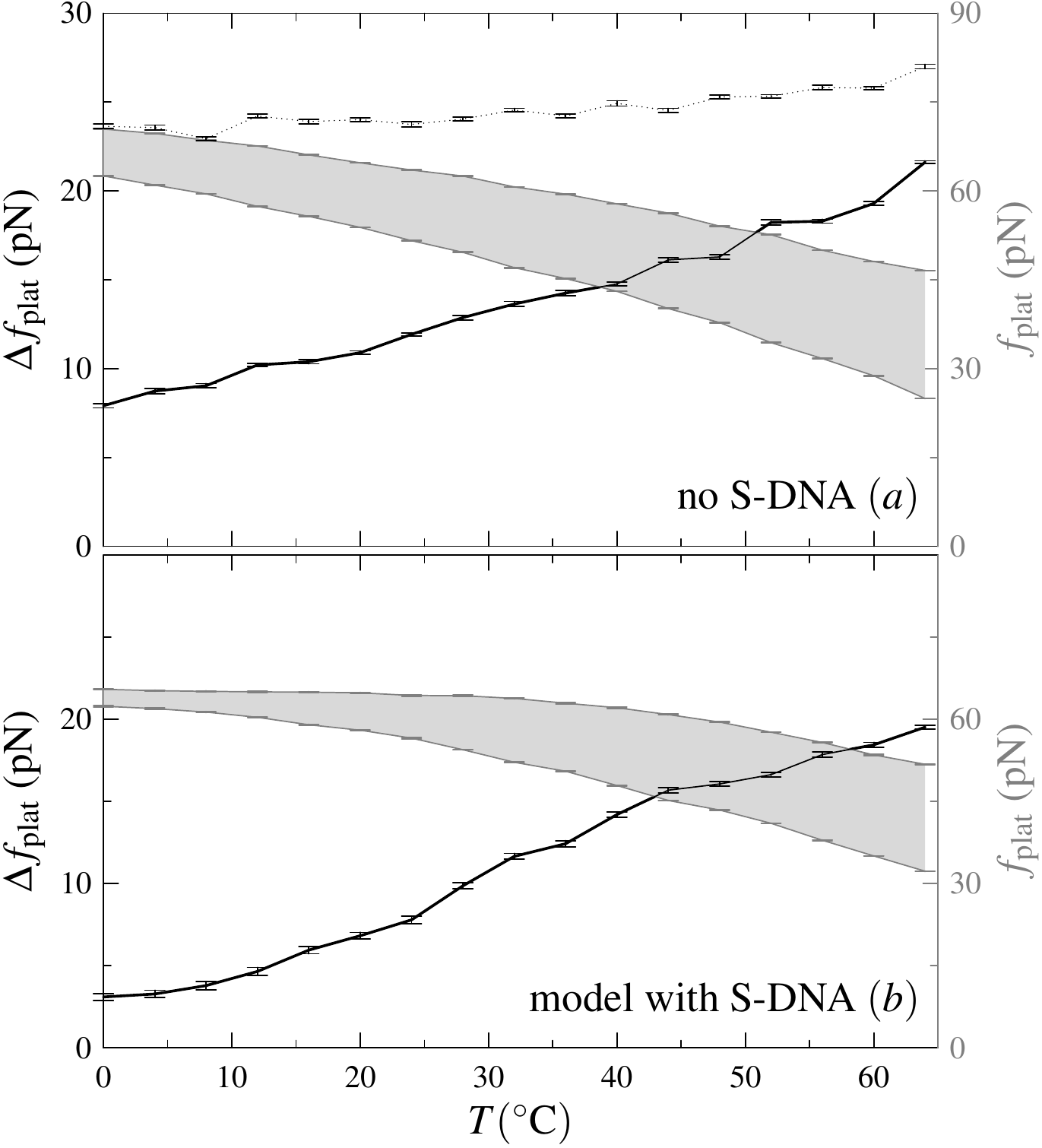} 
   \caption{\label{fig3}Temperature dependence of plateaux height difference $\Delta f_{\rm plat}$ from simulations of 10 kbp chimeric sequences at 150 mM NaCl, for $d=0$ [each data point represents mean of 2 simulations] is not a discriminating test of the two pictures of overstretching. (a) When S-DNA is assumed not to exist the temperature variation of $\Delta f_{\rm plat}$ is governed by the temperature dependence of the specific heat of melting of AT- and CG basepairs. This variation is weak when specific heats are taken to be zero  at (i.e. Taylor expanded about) the melting temperature of $\lambda$-DNA (dashed line), and strong when specific heats are taken to be zero at respective AT- and CG melting temperatures, a more plausible model (solid line). (b) When S-DNA is included in the model the AT-rich half unpeels at all temperatures considered, while the CG-rich half unpeels at high temperature and elongates by way of the B-to-S transition at low temperature. Despite this change in mechanism upon allowing S-DNA, the resulting plateaux height difference varies with temperature in a manner that is qualitatively similar to the solid line in the top panel.}
   \end{figure}
   
We expect that in experiment the effects of secondary structure will play only a minor role in the scenarios we have discussed. While it is likely that hairpins, for instance, will form in separated strands that are very AT- or CG-rich, and while indeed hairpins give rise to additional plateaux at $\sim$ 9 pN (AT) and $\sim$ 20 pN (CG) upon repeated extension~\cite{rief}, we argue that hairpin formation will not influence strongly the heights of the plateaux measured upon {\em first} extension of the DNA. Plateaux heights will be influenced if the dynamics of the unpeeled `front' is restricted by the formation of a hairpin in the non tension-bearing strand. This would be a concern were the front to move diffusively, in which case its reverse movement might be blocked by a hairpin formed in its wake. However, at the pulling rates we consider the motion of the front is super-diffusive, and its dynamics is not strongly affected by potential blockages in its wake.

We end with the observation that varying temperature provides a much less clear assessment of the nature of distinct overstretching transitions than does varying pulling rate. In our simulations, force-extension data for strongly heterogeneous sequences display double plateaux, but the variation of the heights of these plateaux with temperature does not indicate clearly the nature of the responsible overstretching mechanism. We demonstrate this point in Figure~\ref{fig3}. Accounting for the sequence dependence of specific heats of melting, the temperature dependence of the plateaux height difference is {\em not} changed qualitatively by allowing or suppressing the S-state. To make predictions on the basis of the slight difference that does exist would require precise knowledge of the sequence dependence of the specific heat of melting~\cite{bloomfield2}. Indeed, the variation with $T$ of the height of the unpeeling plateaux (accessible even within the B-to-S model at sufficiently high temperature) should allow one to use stretching experiments to {\em measure} the specific heat of melting per basepair. We conclude that the most striking test for S-DNA is kinetic in nature. 

We have shown that within a statistical mechanical model of DNA overstretching one can induce spatial segregation of competing transitions through appropriate sequence design. The force-extension signatures of these distinct transitions vary with pulling rate in a manner that depends qualitatively on the accessibility of an elongated, hybridized state. We propose that the corresponding experiments would provide a means of assessing whether indeed such a state exists. Our results also suggest that de-hybridization of specific locations within a DNA molecule can be effected by subjecting appropriately designed sequences to external force. Such precision melting would permit, for instance, doping of DNA at specific locations, allowing fine control of its conductance properties.
   
We thank J. Ricardo Arias-Gonz\'alez for discussions. SW was supported initially by the DOE, and subsequently by BioSim European Union Network of Excellence (LSHB-CT-2004-005137). SP was supported by the California Institute for Quantitative Biosciences and the NSF. PLG acknowledges support from the Biomolecular Materials Program administered by the Materials Sciences Division of Lawrence Berkeley National Laboratory (DOE grant number DE-AC0205CH11231). SW acknowledges a Royal Society conference grant that made possible a collaborative visit. Computing facilities were provided in part by the Centre for Scientific Computing at the University of Warwick with support from the Science Research Investment Fund. 

\section{Appendix A: Multi-stage elongation of chimeric sequences}

\begin{figure}[h!] 
\label{}
\centering
\includegraphics[width=\linewidth]{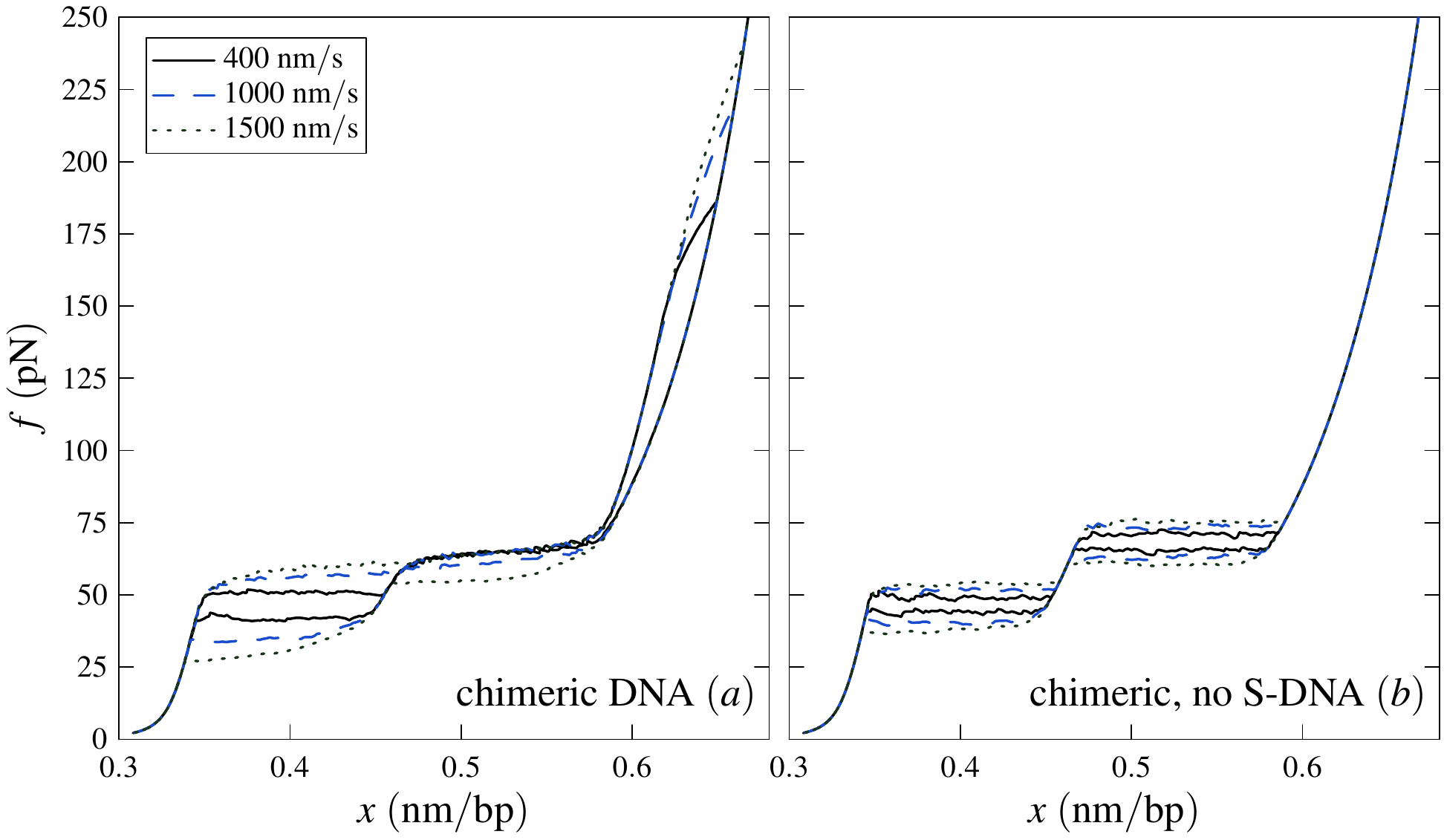} 
 \caption{\label{app1} Overstretching designed sequences to high forces. Simulated chimeric DNA (10 kbp, 150 mM NaCl, 16$\dc$, $d=0$, pullrates 400, 1000 and 1500 nm/s) shows a three-stage transition if S-DNA is included in the model (a). The lowest plateau signals unpeeling of the AT rich half, while subsequent plateaux indicate elongation of the CG-rich half by way of the B-to-S transition, followed at higher force by unpeeling of the same half. If S-DNA is not included in the model (b) then only two transitions are seen. These correspond to unpeeling of AT- and CG-rich halves, respectively.}
 \end{figure}

Simulations performed on chimeric sequences in Figure 1 of the main text (panels b,c) used a maximum pulling force of 120 pN, and reveal two distinct plateaux. At higher forces we observe within the B-to-S model a third plateaux corresponding to a pulling rate-dependent unpeeling of the CG-rich half. A similar unpeeling is seen when stretching $\lambda$-DNA to high forces (see Fig 1(a) and Refs.~\cite{cs,rief,cocco}). We illustrate the multi-stage elongation of chimeric sequences in Figure~\ref{app1} (a): the lowest plateau represents unpeeling of the AT-rich half; the middle plateau signals elongation of the CG-rich half by way of the B-to-S transition; and the highest plateau indicates the subsequent unpeeling of the CG-rich half. At pulling rates considered, transitions signaling unpeeling are pulling rate-dependent, while B-to-S conversion is not. In Figure 1(b) we show for comparison calculations performed with S-DNA not included in the model. In this case the two plateaux signal unpeeling of AT- and CG-rich halves; there is no further transition at higher force.

\begin{figure}[!h] 
\label{}
\centering
\includegraphics[width=\linewidth]{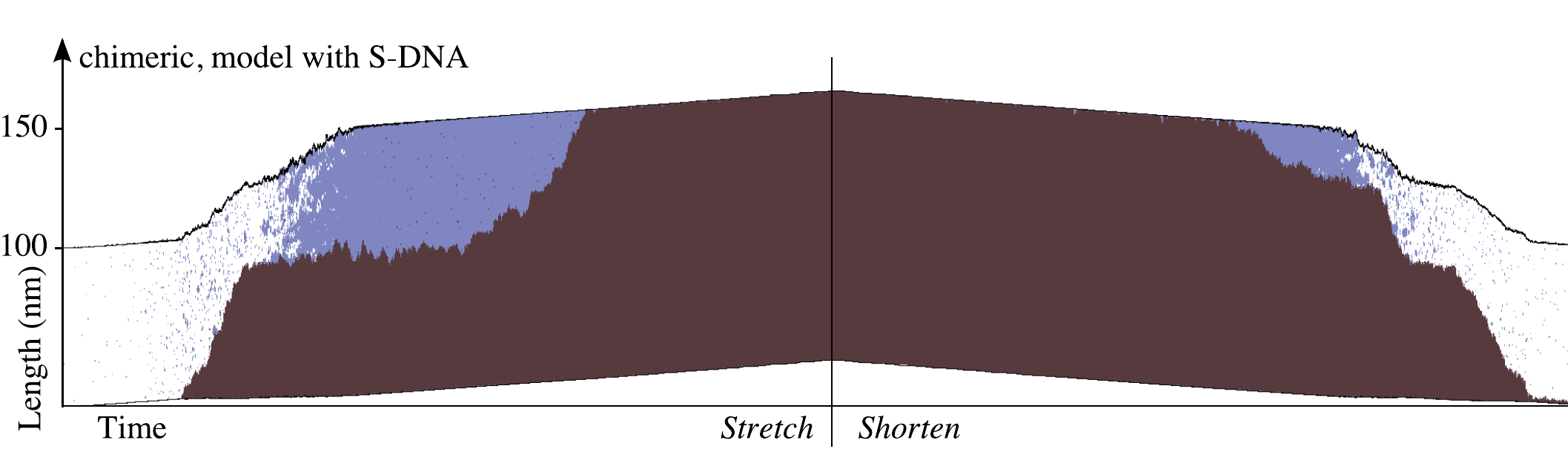} 
\caption{\label{figtraj2}  Microscopic configurations as a function of time from simulations of 300-basepair fragments of chimeric DNA at $21^{\circ}$C and 150 mM NaCl, for pulling rate 1000 nm/s. S-DNA is included in the model. White indicates B-DNA, red indicates unhybridized DNA, and blue indicates S-DNA. The three-stage elongation evident in the force-extension data of Figure~\ref{app1} (a) is seen clearly as a unpeeling of the AT-rich (bottom) half of the molecule, followed by a conversion of the CG-rich (upper) half first to S-DNA and then to unpeeled DNA. The spatial scale indicates molecular length, with the extension modeling the departure of the `trapped' bead from the trap center reduced by a factor of $10$ for clarity. The maximum force attained is 200 pN.}
\end{figure}
In Figure~\ref{figtraj2} we show the microscopic dynamics that underlie the three-stage elongation shown in Figure~\ref{app1} (a). 

\section{Appendix B: Choice of model timescale}

The pulling rate-dependence of unpeeling within our model is governed by the imposed pulling timescale, controlled by the pulling speed $v_0$, and the fundamental timescale $\Gamma_0^{-1}$ on which basepairs change state. In Ref.~\cite{us} we argue, by comparison with experimental data~\cite{rief,cs}, that a  fluorescence-derived time in the microsecond range ($\Gamma_0^{-1}= 28$ $\mu$s~\cite{bubbles1}) is a more appropriate choice for this fundamental timescale than the 10-100 ns fluctuations identified by NMR experiments. Here we illustrate the effect of changing $\Gamma_0$ by determining analytically, as a function of model parameters, the approximate force at which complete unpeeling occurs when stretching once-nicked DNA. We will compare this estimate with the $\lambda$-DNA overstretching data of Refs.~\cite{cs,rief}. These data display a pulling rate-independent `65 pN' plateau followed at higher force by a rate-dependent transition interpreted as an unpeeling. 

We consider within our model the drift of the unpeeled `front' separating hybridized and unhybridized regions, for a sequence of $N$ basepairs bearing a nick at one end. For simplicity we will assume sequence homogeneity, and model the force-extension profile of DNA using the piecewise linear fit of Cocco et al.~\cite{cocco}. This fit assumes a linear overstretching plateau between 62 and 68 pN of gradient $m_1 \equiv (68-62)/(0.58-0.34) \, {\rm pN}/( {\rm nm} \cdot N)$, and a gradient for forces above 68 pN of magnitude $m_2 \equiv 1600 /0.34 \, {\rm pN}/( {\rm nm} \cdot N)$. We shall assume a constant rate of molecular extension, $L = v_0 t$. The master equation derived from our model for the position with time $t$ of the unpeeled front, $n(t)$ ($0 \leq n \leq N$), may be manipulated to yield the front's drift velocity, $\dot{n}(t)= \Gamma_0 (W^{+}(f) - W^{-}(f))$. Here the $W^{\pm}(f) = [1+\exp(\pm \beta [\Delta g - \delta w(f)]) ]^{-1}$ are Glauber rates. The term $\Delta g$ is the (sequence-averaged) free energy difference between hybrizided and unhybridized DNA at zero force, and $\delta w(f)$ models the free energy of extension of the unhybridized phase relative to the hybrdized phase. We shall assume that unpeeling occurs only over S-form DNA, i.e. that the B-to-S conversion just pre-empts unpeeling. For forces large enough that unpeeling is favorable we have
\begin{figure}[!h] 
\centering
\includegraphics[width=0.7\linewidth]{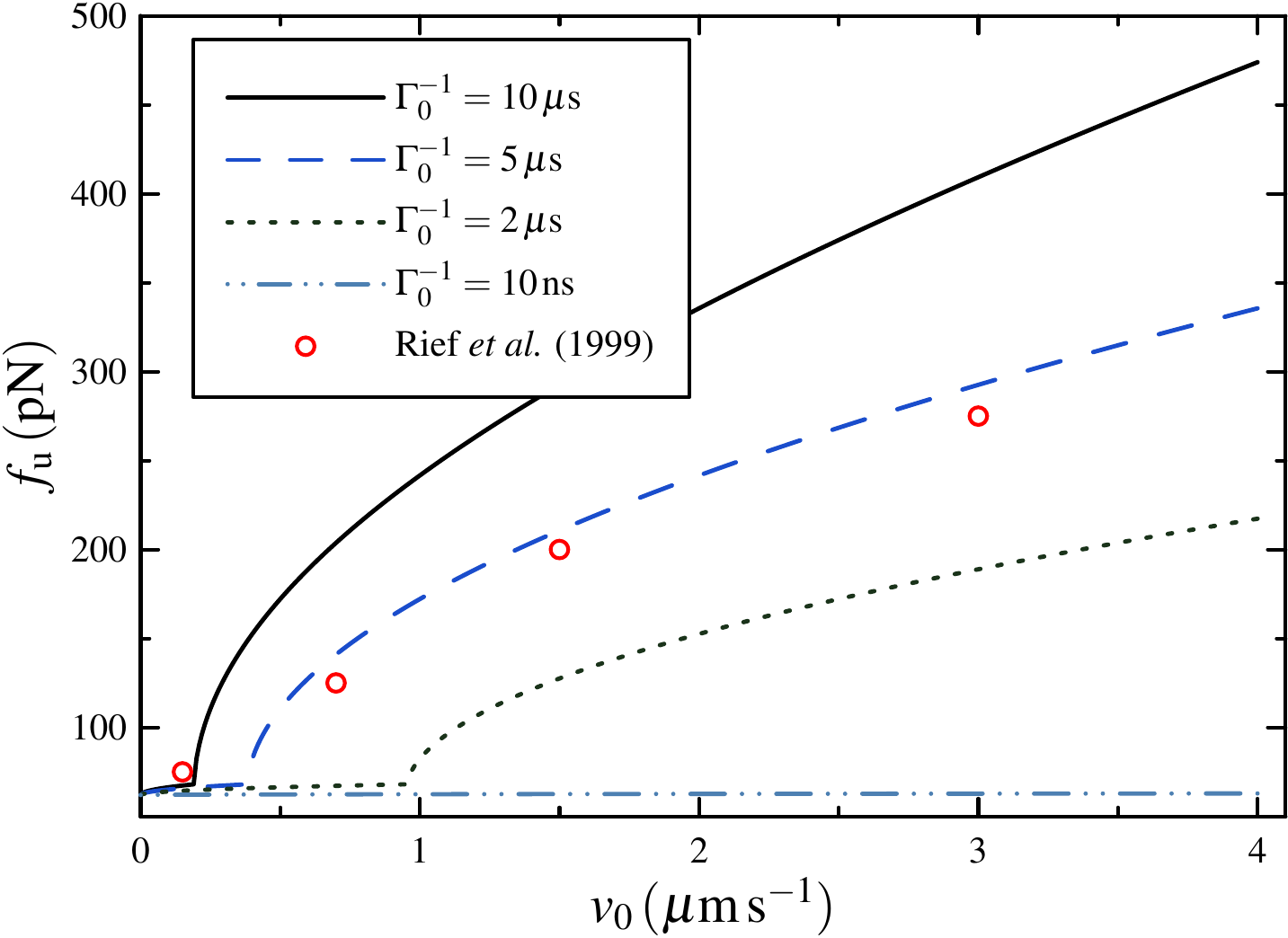} 
 \caption{\label{fig_rief} Approximate unpeeling force $f_{\rm u}$ (derived from Equation~(\ref{peel})) versus pulling speed $v_0$ for sequence-averaged $\lambda$-DNA at a range of model timescales $\Gamma_0^{-1}$. We obtain reasonable comparison with experiment~\cite{rief} for timescales in the microsecond range, characteristic of fluorescence measurements~\cite{bubbles1}.}
 \end{figure}
\beq
\label{eq_peel}
\frac{d n(t)}{dt} \approx \Gamma_0 \frac{ \sinh \beta \hat{f} \Delta L}{1+\cosh \beta \hat{f} \Delta L},
\eeq
where $\Delta L \approx 0.02$ nm is the difference in length per basepair between S- and U-DNA, which we assume to be independent of force (a reasonable approximation near and just above the overstretching force). The variable $\hat{f} \equiv f-f_{\rm eq.}$ is the force in excess of the S-U equilibrium force $f_{\rm eq.}$, which within our model is approximately $62$ pN at 150 mM NaCl and $21^{\circ}$C. We assume that $n(\hat{f}=0)=0$. We change variables in Equation~(\ref{eq_peel}) from $t$ to $L=t/v_0$, and integrate along the piecewise-linear force-extension profile to obtain an expression for $n$ in terms of $\hat{f}$:
\beq
n(\hat{f}) = \frac{\Gamma_0}{v_0 \beta \Delta L} \int_0^{\hat{f}} \left( \frac{d L}{d \hat{f}'} \right) d \hat{f}' \frac{\partial}{\partial \hat{f}'} \ln |1+\cosh \beta \Delta L \hat{f}' |.
\eeq
 This expression may be manipulated to yield
\bea
\label{peel}
{\cal C}( \hat{f})=  \left\{
 \begin{array}{lc}
      k_2^{-1} n-K [{\cal C}(\hat{f}_1)- {\cal C}(0)]+{\cal C}(\hat{f}_1) &(\hat{f} \geq \hat{f}_1) \\
      k_1^{-1} n+{\cal C}(0)& (0 \leq \hat{f} < \hat{f}_1)
       \end{array} .
             \right.
\eea
We have defined ${\cal C}(f) \equiv \ln |1+\cosh \beta \Delta L f |$; $\hat{f}_1 \equiv (68 -62) $ pN $= 6 $ pN; $k_i(v_0) \equiv \Gamma_0/( \beta \Delta L v_0 m_i)$, for $i=1,2$; and $K(v_0) \equiv k_1/k_2$. Recall that $\beta \approx 1/(4.1 \, {\rm pN \, nm})$. Setting $n = N$ in Equation (\ref{peel}) gives the excess force $\hat{f}_{\rm u} = f_{\rm u}-f_{\rm eq.}$ at which complete unpeeling occurs. For low pulling rates this excess force scales as $ \beta \Delta L \hat{f}_{\rm u} \approx 2 \left( \beta \Delta L v_0 m_1 N \Gamma_0^{-1} \right)^{1/2}$, while for very large pulling rates we have $\beta \Delta L \hat{f}_{\rm u} \sim \beta \Delta L v_0 m_2 N \Gamma_0^{-1}$. We plot the unpeeling force $f_{\rm u}=\hat{f}_{\rm u} + f_{\rm eq.}$ against pulling rate in Figure~\ref{fig_rief} for different choices of the fundamental timescale $\Gamma_0^{-1}$. We include in this plot data extracted from Figure 2 of Ref.~\cite{rief}. We expect that our estimate of $f_{\rm u}(v_0)$ should be accurate to within an order of magnitude; we obtain favorable comparison with experiment for $\Gamma_0^{-1}$ in the microsecond range. The equation of the line in Figure~\ref{fig_rief} calculated with $\Gamma_0^{-1} = 5 \, \mu$s is
\bea
\label{peel2}
f=  62+\left\{
 \begin{array}{lc}
      205 \cosh^{-1} \left(1.915 \, e^{0.1148 v} -1 \right) & (f \geq{\rm 68 \, pN})\\
   205 \cosh^{-1} \left(2 \, e^{0.0000563 v}-1\right)    &( f  < {\rm 68 \, pN}) 
       \end{array} .
             \right.
\eea
\bibliography{bib}
\end{document}